\documentstyle[12pt]{article}
\def\Msun{\ifmmode{~M_\odot}\else$M_\odot$~\fi}
\def\kms{\ifmmode{$~km\thinspace s$^{-1}~}\else km\thinspace s$^{-1}~$\fi}
\def\ga{\mathrel{\mathchoice {\vcenter{\offinterlineskip\halign{\hfil
$\displaystyle##$\hfil\cr>\cr\noalign{\vskip1.5pt}\sim\cr}}}
{\vcenter{\offinterlineskip\halign{\hfil$\textstyle##$\hfil\cr>\cr
\noalign{\vskip1.0pt}\sim\cr}}}
{\vcenter{\offinterlineskip\halign{\hfil$\scriptstyle##$\hfil\cr>\cr
\noalign{\vskip0.5pt}\sim\cr}}}
{\vcenter{\offinterlineskip\halign{\hfil$\scriptscriptstyle##$\hfil
\cr>\cr\noalign{\vskip0.5pt}\sim\cr}}}}}
\def\la{\mathrel{\mathchoice {\vcenter{\offinterlineskip\halign{\hfil
$\displaystyle##$\hfil\cr<\cr\noalign{\vskip1.5pt}\sim\cr}}}
{\vcenter{\offinterlineskip\halign{\hfil$\textstyle##$\hfil\cr<\cr
\noalign{\vskip1.0pt}\sim\cr}}}
{\vcenter{\offinterlineskip\halign{\hfil$\scriptstyle##$\hfil\cr<\cr
\noalign{\vskip0.5pt}\sim\cr}}}
{\vcenter{\offinterlineskip\halign{\hfil$\scriptscriptstyle##$\hfil
\cr<\cr\noalign{\vskip0.5pt}\sim\cr}}}}}
\def\appendix{\par
 \setcounter{section}{0}
 \def\thesection{A\arabic{section}}
 \setcounter{equation}{0}
 \def\theequation{A\arabic{equation}}
 \setcounter{figure}{0}
 \def\thefigure{A\@arabic\c@figure}
 \setcounter{table}{0}
 \def\thetable{A\@arabic\c@table}
}

\begin{document}

\title{{\bf On the global structure of self-gravitating disks for softened gravity}}
\author{\\
Jesper Sommer-Larsen$^{1}$,
Henrik Vedel$^{1}$ and
\and Uffe Hellsten$^{2}$\\
\\
$^1$Theoretical Astrophysics Center\\
Juliane Maries Vej 30, DK-2100 Copenhagen {\O}, Denmark\\
(jslarsen@tac.dk, vedel@tac.dk)\\
\\
$^2$University of California, Lick Observatory,\\
Santa Cruz, CA 95064, USA\\
(uffe@ucolick.org)\\
\\}
\date{}
\maketitle


\section*{Abstract}
Effects of gravitational softening on the global structure of self-gravitating
disks in centrifugal equillibrium are examined in relation to 
hydrodynamical/gravitational simulations. The one-parameter spline softening
proposed by Hernquist \& Katz is used.

It is found that if the characteristic size of a disk, $r$, is comparable to or less than
the gravitational softening length, $\epsilon$, then the cross section of the\\
\\
{\it Submitted to Monthly Notices of the Royal Astronomical Society}\\
\\
simulated disk is significantly larger than that of a no-softening (Newtonian) disk with the
same mass and angular momentum.

We furthermore demonstrate that if $r \la \epsilon/2$ then the scaling relation
$r \propto \epsilon^{3/4}$ holds for a given mass and specific angular momentum 
distribution with mass.
Finally we compare some of the theoretical results obtained in this and a previous
paper with the results of numerical Tree-SPH simulations and find qualitative 
agreement.

\vskip 0.9 truecm {\bf Key words:} ~galaxies: kinematics and dynamics --
~galaxies: structure -- ~methods: numerical
\newpage

\section{Introduction}
In pure N-body as well as gravitational/hydrodynamical, particle-based simulations
the gravitational field of the individual particles is commonly softened, primarily
to suppress effects of two-body gravitational interactions - see, e.g., Evrard (1988), 
Hernquist \& Katz (1989), 
Sommer-Larsen, Vedel \& Hellsten (1996, paper I in the following).

In paper I we demonstrated that one has to be cautious when comparing
results of such simulations
with reality when gravity has been softened. This was illustrated by considering,
as an example, non-rotating, self-gravitating, isothermal spheres in hydrostatic equillibrium
showing that by introducing gravitational softening the structure
of such isothermal spheres can be dramatically changed relative to the Newtonian
case. This occurs, in qualitative terms, when the radial extent of the spheres is less
than or of the order the gravitational softening length, $\epsilon$ (in paper I as well
as in this paper the one-parameter spline softening proposed by Hernquist \& Katz (1989)
is used).

It also follows from paper I that for isothermal spheres of a given mass 
and temperature the size increases with increasing $\epsilon$. In particular, for 
isothermal spheres of characteristic scale $r \la \epsilon/2$ (and given mass and
temperature) it follows from paper I that $r \propto \epsilon^{3/2}$.

In this paper we study the global structure of self-gravitating disks in centrifugal
equillibrium for softened gravity. We show that if the characteristic size $r$ of the disks
is less than or of the order $\epsilon$, then the size of the disks is significantly larger
than for the Newtonian case for a given mass and angular momentum. We also
show that if the characteristic size of the disk $r \la \epsilon/2$
then $r \propto \epsilon^{3/4}$ (for a given mass and specific angular momentum 
distribution with mass).

In section 2 we describe the effects of gravitational softening on the global structure
of self-gravitating disks and discuss the results obtained. Finally section 3 
constitutes the conclusion.

\section{Effects of gravitational softening on the global structure of self-gravitating disks 
in centrifugal equillibrium}
The surface density of a stellar, galactic disk is in general well approximated by a
truncated exponential (e.g. Freeman 1970, Van der Kruit 1987)
\begin{equation} \Sigma(R) = \left \{ \begin{array}{ll}
                                                    \Sigma_0 \exp (-R/R_d) ~,  & ~~R < R_t \\
                                                    0 ~,                                                     & ~~R \ge R_t 
                                                  \end{array}
                                 \right. 
\end{equation}
where $\Sigma_0$ is the central surface density, $R_d$ the exponential
scale length, $R_t$ the truncation radius ($R_t \sim 4 R_d$) and $R$ the radial
coordinate in the plane of the disk.
Consider first, for a given mass and angular momentum, the one-parameter
family of self-gravitating, infinitely flat disks in centrifugal equillibrium described
by eq. [1] (with $R_t = 4 R_d$) for various values of the gravitational
softening length $\epsilon$. As $\epsilon$ increases and the gravitational field of
the disk becomes increasingly softened one would expect $R_d$ to increase
(for a given mass and angular momentum) and
hence the characteristic cross section of the disk
\begin{equation}
\sigma_c = \pi R_{rms}^2, ~~{\rm where} ~~R_{rms} \equiv <R^2>^{1/2} ~~,
\end{equation}
compared to the Newtonian gravity value $\sigma_{c,0}$ - in eq. [2] $<R^2>$
denotes the mass weighted average of $R^2$. In
Figure 1 $\sigma_c/\sigma_{c,0}$ is shown as a function of the characteristic
size of the disk $R_{rms}$ in units of $\epsilon$. As can be seen from the figure,
when the characteristic size of the disk is equal to the gravitational softening length, 
$\sigma_c$ is about five times larger than $\sigma_{c,0}$ due to the 
effect of gravitational softening. For $R_{rms} \simeq 0.4 \epsilon$ this ratio has
increased further to about two orders of magnitude. In the limit $R_{rms} \ll 
\epsilon$, $\sigma_c/\sigma_{c,0} \propto (R_{rms}/\epsilon)^{-6}$, as can be
easily verified.

Clearly, gravitational/hydrodynamical simulations where self-gravitating disks
of size comparable to or smaller than the gravitational softening length occur have
to be interpreted with care.

In the simple calculations above the mass and angular momentum was assumed
fixed, but clearly the specific angular momentum distribution, $j(M)$, 
where $M$ is the cumulative mass within radius $R$,
changes with $\epsilon$. Indeed, for Newtonian gravity and a given self-gravitating
disk in centrifugal equillibrium with surface density
$\Sigma(R;\epsilon$=0) and corresponding specific angular momentum distribution
$j(M)$, 
a disk solution $\Sigma(R;\epsilon)$ with
the same mass and $j(M)$ does not necessarily exist for softened gravity with a given
non-zero softening length $\epsilon$.

For a disk with surface density given by eq. [1] for Newtonian gravity
($\epsilon=0$, $\Sigma_0 = R_d = 1$, $R_t = 4 R_d$) we were able to obtain disk solutions 
$\Sigma(R;\epsilon)$
with the same mass and $j(M)$ for $\epsilon \ga R_d$ using a numerical,
iterative algorithm. The solutions are shown in Figure 2 as $\epsilon^{3/2} \Sigma$
as a function
of $\epsilon^{-3/4} R$ (see below) for $R_d/\epsilon = 2^{-n}, n=0,2,..,8$.
The corresponding values of $\sigma_c/\sigma_{c,0}$ are plotted in
Figure 1 as filled circles for $n=0,1,..,7$.

For $0<\epsilon\la R_d$ no solutions could be obtained - we comment further on this
below.

\subsection{The linear approximation and self-similarity}
By an argument similar to the one given in paper I it can be shown
that for a system of linear size less than about half
the gravitational softening length the gravitational acceleration at
$R \la \epsilon/2$ is approximately linear in R:
\begin{equation}
g_\epsilon(R) = -q R + O(R^3),\;\; {\rm for} ~~R \la \epsilon/2 \;\;,
\end{equation}
where
\begin{equation}
q= \frac{4 G M_d}{3 \epsilon^3} \;\;,
\end{equation}
$G$ being the gravitational constant and $M_d$ the total mass of the disk.
In this limit it is possible to obtain a disk solution $\Sigma(R;\epsilon)$
given a disk $\Sigma(R;0)$ with a corresponding $j(M)$:

By definition
\begin{equation}
j(M) = v_c(R) R \;\;,
\end{equation}
where the circular speed at $R$ is given by (Binney \& Tremaine 1987)
\begin{equation}
v_c(R) = R^{1/2} \left [2 \pi G \int_0^\infty [\int_0^\infty  J_0(kR')
\Sigma(R';0) R' dR'] ~ J_1(kR) k dk \right ]^{1/2} \;\;,
\end{equation}
where $J_0$ and $J_1$ $ (= -J_0')$ are Bessel functions of the first kind, and
\begin{equation}
M(R) = 2 \pi \int_0^R \Sigma(R';0) R' dR' \;\;.
\end{equation}
For later use
\begin{equation}
\frac{dj}{dM} = \frac{dj}{dR} \frac{dR}{dM} =  
\frac{1}{2 \pi R \Sigma(R;0)} [v_c(R) + R\frac{dv_c(R)}{dR}] \;\;.
\end{equation}

Now, in the linear approximation (assuming $\epsilon \ne 0$)
\begin{equation}
g_\epsilon(R) = -q R, \;\; \Rightarrow \;\; v_c(R) = q^{1/2} R \;\;,
\end{equation}
so 
\begin{equation}
j(R) = q^{1/2} R^2 \;\;,
\end{equation}
and
\begin{equation}
\frac{dj}{dM} = \frac{dj}{dR} \frac{dR}{dM} =  
\frac{2 q^{1/2} R}{2 \pi R \Sigma(R;\epsilon)} = \frac{q^{1/2}} 
{\pi \Sigma(R;\epsilon)} \;\;.
\end{equation}
Hence
\begin{equation}
\Sigma(R;\epsilon) = (\frac{4 G M_d}{3 \pi^2})^{1/2}\;
\epsilon^{-3/2} (\frac{dj}{dM})^{-1} \;\;,
\end{equation}
where, by eq. [10],
\begin{equation}
R(j) = q^{-1/4} j^{1/2} = (\frac{3}{4 G M_d})^{1/4} \epsilon^{3/4} j^{1/2} ~~.
\end{equation}
Given a $\Sigma(R;0)$ eqs. [6]-[8], [12] and [13] can be used to
determine the corresponding $\Sigma(R;\epsilon)$. It is easy to show that
the solutions are self-similar:
\begin{equation}
\Sigma(R;\alpha\epsilon) = \alpha^{-3/2}\;\Sigma(\alpha^{-3/4}R;\epsilon) 
\;\;.
\end{equation}

For $\Sigma(R;0)$ given by eq. [1], $\epsilon^{3/2} \Sigma(R;\epsilon)$ 
as a function of $\epsilon^{-3/4} R$ is shown in
Figure 2 as a solid line. The solutions obtained previously for
$R_d/\epsilon = 2^{-n}, n=0,2,..,8$, as displayed in Figure 2, converge 
towards this limiting solution as $R_d/\epsilon \rightarrow 0$, as expected.

Given that in the linear approximation a solution $\Sigma(R;\epsilon)$ with the
same mass and specific angular momentum distribution as the Newtonian disk 
$\Sigma(R;0)$ can be obtained (eq. [12]) it is not surprising that such solutions
in general can be obtained for $R_d/\epsilon$ less than some critical value. But,
as described in the previous subsection, one should not expect to be able to obtain
such solutions for all values of $R_d/\epsilon$.

In the simplified Tree-SPH galaxy formation simulations described in paper I
self-gravitating, quasi-isothermal gas systems of characteristic 
size $r \la \epsilon/2$ occur
at $t$ = 3.2 Gyr in three, otherwise identical, simulations with $\epsilon$ =
1.5, 3.0 and 6.0 kpc respectively. As discussed in paper I the systems become
increasingly pressure supported relative to rotational support as $\epsilon$
increases.

For systems of equal mass and temperature it follows from eqs. [29] and [31]
of paper I that one would expect $r \propto \epsilon^{3/2}$ for pressure supported 
systems. Likewise it follows from eq. [14] above that for rotationally supported systems 
of equal mass and $j(M)$ one would expect $r \propto \epsilon^{3/4}$. In
Figure 3 we have plotted the half-mass radii of the most massive, self-gravitating
system at $t$ = 3.2 Gyr in the three simulations as a function of $\epsilon$.
As can be seen from this Figure $r_{1/2}(\epsilon)$ increases approximately as
$\epsilon^{3/4}$ as $\epsilon$ increases from 1.5 to 3.0 kpc and  
approximately as  $\epsilon^{3/2}$ as $\epsilon$ increases further from
3.0 to 6.0 kpc in good agreement with the increasing degree of pressure support
found in paper I.

\section{Conclusion}
In hydrodynamical/gravitational simulations involving gravitational softening
an adverse effect of the softening is to make simulated, self-gravitating disks
artificially large by a significant factor if the characteristic size, $r$, of the simulated
disks is less than or comparable to the gravitational softening length, $\epsilon$. In
such cases the outcome of the simulations should be interpreted with due care.

If $r \la \epsilon/2$ then the scaling relation $r \propto \epsilon^{3/4}$ holds for
disks with a given mass and specific angular momentum distribution $j(M)$.

Finally, the theoretical results obtained in this paper and paper I are compared
with the results of numerical, Tree-SPH simulations and qualitative agreement
is found.

\section*{Acknowledgements}
This work was supported by Danmarks Grundforskningsfond through its support
for an establishment of the Theoretical Astrophysics Center.
UH acknowledges support by a postdoctoral research grant from the
Danish Natural Science Research Council.

\section*{References}
\begin{trivlist}
\item[] Binney, J., Tremaine, S., 1987, Galactic Dynamics. Princeton Univ.
Press,
\item[] \hspace{1.0cm} Princeton
\item[] Evrard, A.E., 1988, MNRAS, 235, 911
\item[] Freeman, K.C., 1970, ApJ, 160, 811
\item[] Hernquist, L., Katz, N., 1989, ApJS, 70, 419 
\item[] Sommer-Larsen, J., Vedel, H., Hellsten, U., 1996, ApJ, submit. (paper I)
\item[] Van der Kruit, P.C., 1987, A\&A, 173, 59
\end{trivlist}

\newpage
\section*{Figure captions}
Figure 1:  The ratio of the characteristic cross section of an exponential disk for
softened gravity to that of a Newtonian exponential disk for a given mass and
angular momentum (solid line). The dots show the same quantities for solutions
of given mass and specific angular momentum distribution $j(M)$.\\
\\
Figure 2:  The surface density distribution in the linear approximation of a disk of 
the same mass and specific angular momentum distribution, $j(M)$, 
as a Newtonian,
exponential disk with $R_t = 4 R_d$, displayed as $\epsilon^{3/2} \Sigma$ vs.
$\epsilon^{-3/4} R$
 (solid line).  Other curves: Solutions of the
same mass and $j(M)$ for $R_d/\epsilon = 2^{-n}, n=0,2,..,8$ also displayed as 
$\epsilon^{3/2} \Sigma$ vs. $\epsilon^{-3/4} R.$
The solutions converge
monotonically toward the limiting solution shown by the solid line as $n$ increases.\\
\\
Figure 3: Half-mass radius as a function of $\epsilon$ for the most massive, dense
gas clumps at $t = 3.2$ Gyr occurring in the simulations described in paper I. Also
shown are power law relations of logarithmic slope 3/4 and 3/2 respectively.\\

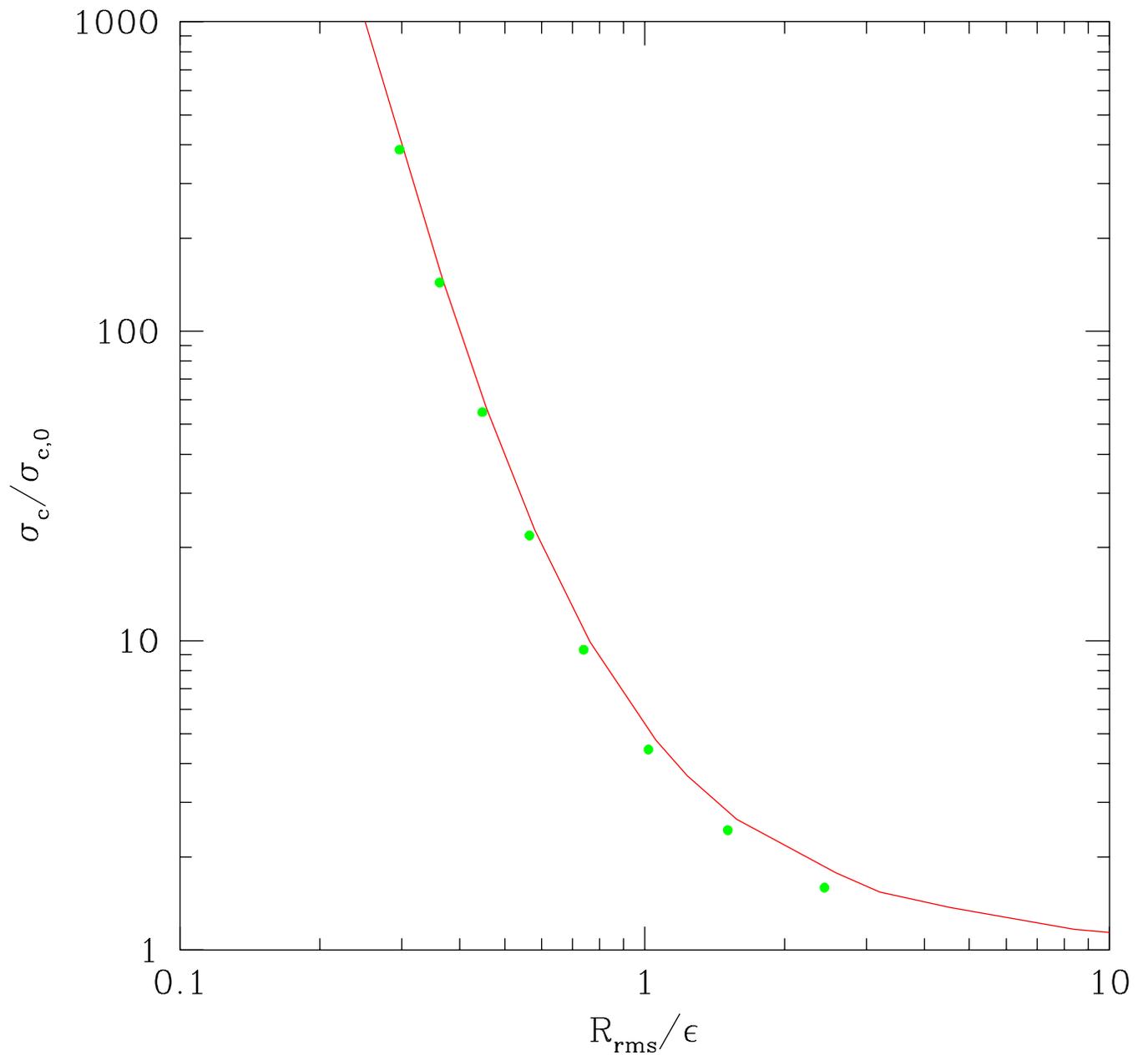
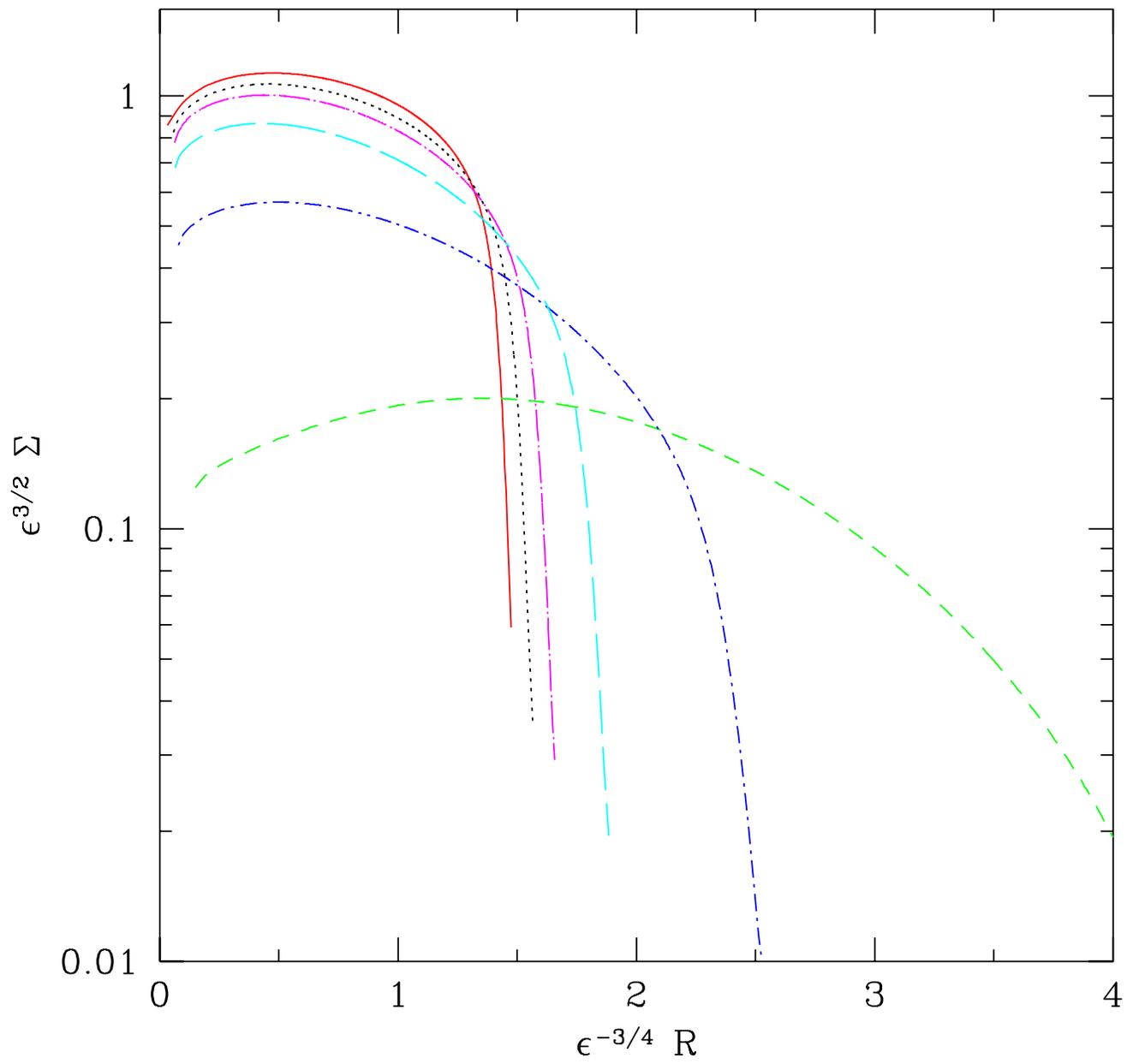
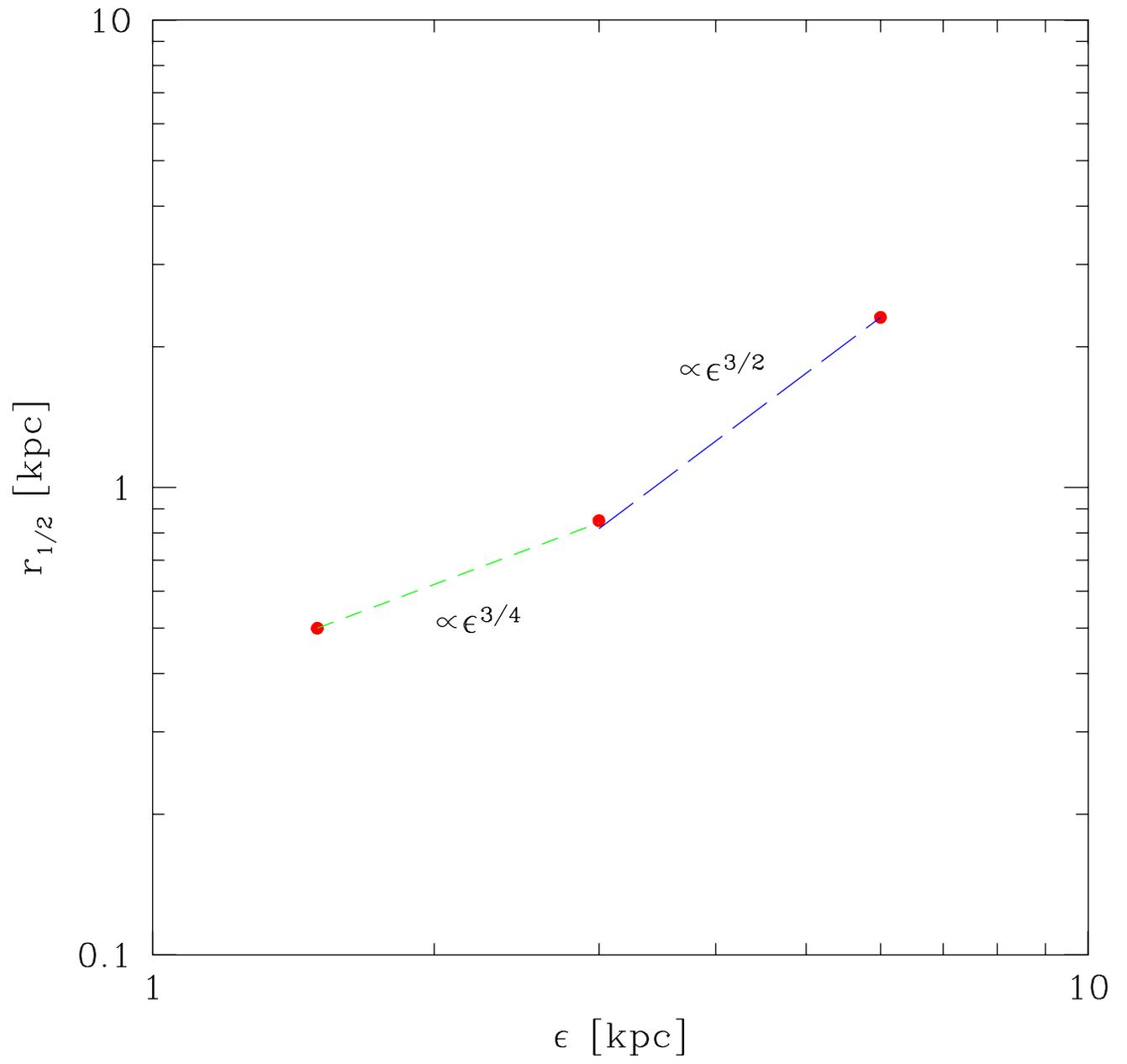
\end{document}